\documentclass{iau}
\usepackage{graphicx}

\title
{It takes a supercluster to raise a galaxy}

\author[Lietzen \& Einasto]   
{Heidi Lietzen$^{1,2}$
 \and Maret Einasto$^3$}

\affiliation{$^1$Instituto de Astrof\'isica de Canarias, E-38205 La Laguna, Tenerife, Spain \\[\affilskip]
$^2$Universidad de La Laguna, Dept. Astrof\'isica, E-38206 La Laguna, Tenerife, Spain \\email: {\tt hlietzen@iac.es}\\[\affilskip]
$^3$Tartu Observatory, 61602 T\~oravere, Estonia }

\pubyear{2014}
\volume{308}  
\pagerange{??--??}
\date{?? and in revised form ??}
\jname{The Zeldovich Universe: Genesis and Growth of the Cosmic Web}
\editors{Rien van de Weygaert, Sergei Shandarin, Enn Saar \& Jaan Einasto, eds.}
\begin{document}

\maketitle

\begin{abstract}
The properties of galaxies depend on their environment: red, passive elliptical galaxies are usually located in denser environments than blue, star-forming spiral galaxies. This difference in galaxy populations can be detected at all scales from groups of galaxies to superclusters. In this paper, we will discuss the effect of the large-scale environment on galaxies. Our results suggest that galaxies in superclusters are more likely to be passive than galaxies in voids even when they belong to groups with the same richness. In addition, the galaxies in superclusters are also affected by the morphology of the supercluster: filament-type superclusters contain relatively more red, passive galaxies than spider-type superclusters. These results suggest that the evolution of a galaxy is not determined by its local environment alone, but the large-scale environment also affects.
\end{abstract}

\firstsection 
\section{Introduction}

Properties of galaxies depend on their environment. In dense environments, massive galaxies are more likely to be red and passive ellipticals than in less dense regions. This color-density or morphology-density relation can be found on many scales of density from the distance to the neighboring galaxies to the location of the galaxy in the cosmic web (\cite{Dressler1980,Postman1984,Gomez2003,Einasto1987,Balogh2004,Porter2008,Skibba2009}).

The dependence of galaxy properties on the environment can be explained with an evolutionary scenario, where the star formation in a galaxy is quenched by some interaction with the environment. There are several physical processes that can cause the transformation of galaxies from star forming to passive. These include interaction with other galaxies such as mergers (\cite{Barnes1992}) and galaxy harassment (\cite{Richstone1976,Moore1996}), and interaction with the group or the cluster, such as tidal stripping (\cite{Gnedin2003}), ram pressure stripping (\cite{Gunn1972}), or strangulation (\cite{Larson1980}).

\begin{figure}[b]
\begin{center}
 \includegraphics[width=3.4in]{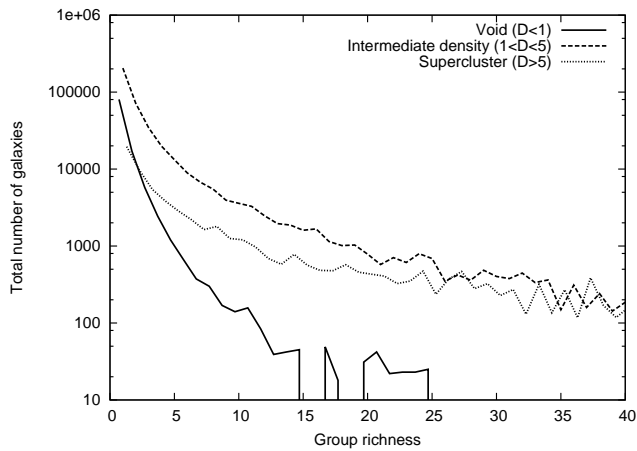} 
 \caption{Distribution of galaxies in voids (solid line), intermediate densities (dashed line), and superclusters (dotted line).}
   \label{fig1}
\end{center}
\end{figure}

We analyse the galaxy populations in different large-scale environments, concentrating especially on galaxies in small groups. The majority of galaxies are located in groups of a few galaxies or in the field, i.e. not belonging to any group. Figure~\ref{fig1} shows the total number of galaxies in different large-scale environments as a function of group richness. The data is from the Sloan Digital Sky Survey 10th Data Release (SDSS DR10). The group catalog was constructed by \cite{Tempel2014} and it contains 588\,193 galaxies that form 82\,458 groups. The sample is magnitude limited with $r<17.77$. The large-scale environment was determined using a luminosity-density field with smoothing scale of 8~Mpc (\cite{Tempel2014,Liivamagi2012}). The unit for the large-scale density $D$ is the mean density of the whole field. Regions with less than the mean density can be defined as voids, while densities with more than five times the mean density are superclusters.

The high fraction of galaxies in groups suggests that the groups are important environments in galaxy evolution. Small groups can also be found in any large-scale environment from voids to superclusters, which makes it possible to compare the different large-scale environments. Our aim is to study how the large-scale environment affects the galaxies. We compare the properties of galaxies in high and low density large-scale environments (superclusters and voids), concentrating especially on galaxies in small groups.

\section{Results}

In \cite{Lietzen2012} we used the galaxy catalogs of the eigth data release (DR8) of the SDSS to compare galaxy populations in different large-scale environments. We found that galaxies in superclusters are more likely to be red and passive than galaxies in voids even when they belong to groups with the same richness. This difference was found by comparing the fractions of star-forming and passive galaxies, and by calculating the average colors of galaxies in different environments. Morphology on its own does not depend on the environment: environments of non-star-forming spiral galaxies are similar to those of elliptical galaxies.

\begin{figure}[b]
\begin{center}
 \includegraphics[width=3.4in]{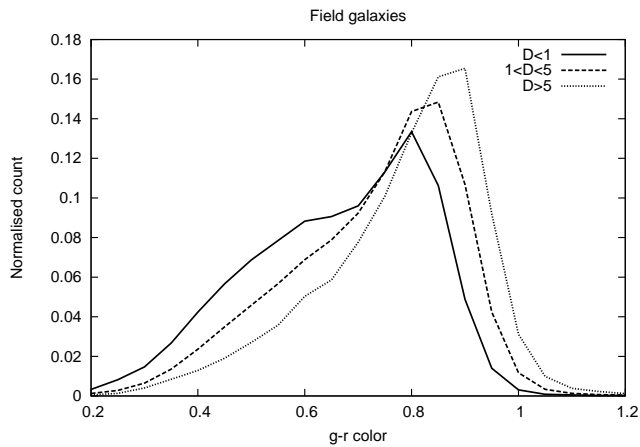} 
 \caption{Color distribution of galaxies in voids (solid line), intermediate densities (dashed line), and superclusters (dotted line).}
   \label{fig2}
\end{center}
\end{figure}

As an example of the large-scale effect, we analyse in Fig.~\ref{fig2} the color distribution of galaxies that do not belong to any group (field galaxies) from the SDSS tenth data release (DR10) data. These galaxies represent the poorest group-scale environments in the universe. The distribution is shown separately for supercluster, intermediate density, and void galaxies. The figure shows that field galaxies in superclusters are more likely red than the field galaxies in voids.

\begin{figure}[b]
\begin{center}
\includegraphics[width=12.0cm]{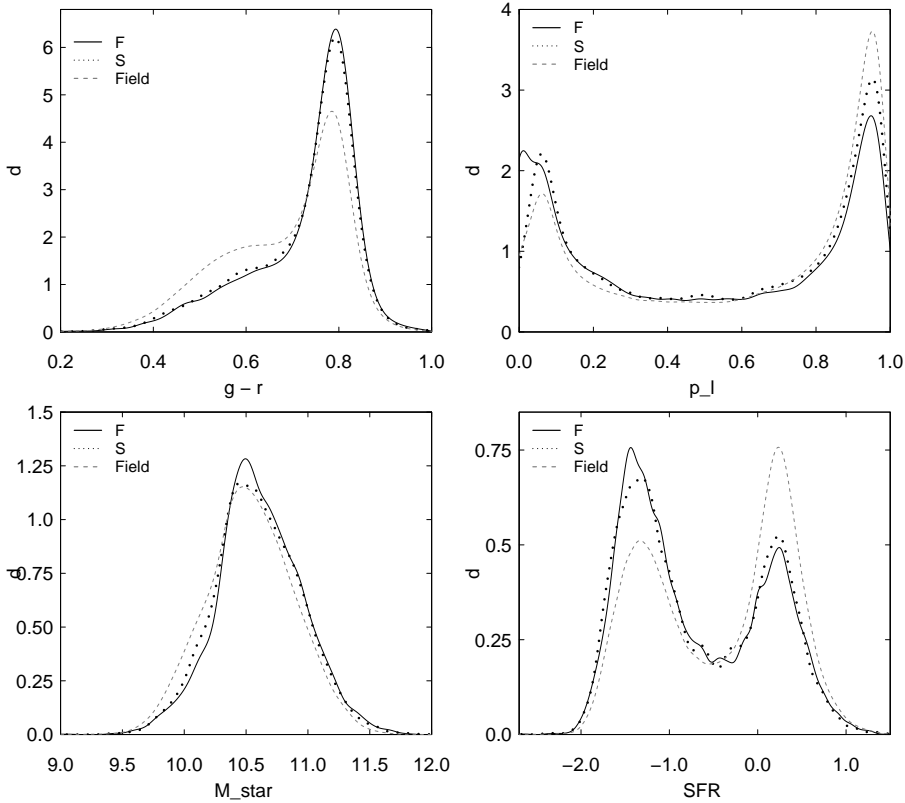} 
 \caption{Distribution of $g-r$ color (top left), probability for late-type morphology (top right), stellar mass (bottom left), and star-formation rate (bottom right) for galaxies in filament (solid line) and spider (dotted line) type superclusters and outside superclusters (dashed line).}
   \label{fig3}
\end{center}
\end{figure}

Besides the differences between superclusters and low-density environments, there are also differences in the galaxy populations between different superclusters. Superclusters can be divided into two morphological types: spider-type superclusters, which consist of high-density clumps with several outgoing filaments, and filament-type superclusters, in which high-density clumps are connected by only a few galaxy filaments. The galaxy populations in different superclusters were studied in \cite{Einasto2014} by comparing galaxy contents of spider and filament type superclusters. Distributions of color, morphology, stellar mass, and star-formation rate of galaxies in the two types of superclusters and in lower-density environments are shown in Fig.~\ref{fig3}. Superclusters of filament type contain a higher fraction of early-type, low star-formation rate galaxies than the spider-type superclusters. There are also significant differences between the galaxy populations of individual superclusters.

\section{Discussion}

Our results suggest that galaxies in superclusters are different from galaxies in voids, and that different superclusters have different galaxy populations. These results imply that the evolution of galaxies may not be determined by the group-scale environment alone, but the large-scale environment also has an effect.

The physical processes that drive galaxy evolution happen on smaller scales. Star formation in a galaxy may be ended by an interaction with a neighboring galaxy or with the surrounding group of galaxies. Typical sizes of superclusters are of the order of ten megaparsec or more. Therefore, the large-scale effect must be less direct. One possible explanation may be the ''speed`` of evolution: according to \cite{Einasto2005}, the dynamical evolution in high-density global environments the dynamical evolution starts earlies and is more rapid than in low-density environments.

The higher fraction of mature galaxies in denser environments may be interpreted as a product of mass assembly bias. \cite{Gao2005} found in the simulations that the oldest 10\,\% of halos are more than five times more strongly correlated than the youngest 10\,\% with the same mass. If old halos also host old galaxies, this could explain the denser environments of the red, passive galaxies.

The morphology of superclusters may affect galaxies in superclusters through different dynamical structure. According to \cite{Einasto2012} superclusters with spider morphology have richer inner structure than those with filament morphology. This means that mergers of clusters may occur more often in the spider superclusters, making them dynamically younger.



\end{document}